\documentclass{article}
\usepackage{waspaa23,amsmath,graphicx,url,times}
\usepackage{color}
\usepackage{flushend}
\usepackage{cite}

\usepackage{ifthen}
\usepackage{amsmath,amssymb,amsfonts,mathtools,nicefrac,bm}
\usepackage{glossaries}
\newacronym[firstplural=spatio-temporal covariance matrices (STCMs)]{STCM}{STCM}{spatio-temporal covariance matrix}
\newacronym{BLCMP}{BLCMP}{binaural linearly constrained minimum power}
\newacronym{RIR}{RIR}{room impulse response}
\newacronym{LCMP}{LCMP}{linearly constrained minimum power}
\newacronym{wBLCMP}{wBLCMP}{weighted binaural linearly constrained minimum power}
\newacronym{STFT}{STFT}{short-time Fourier transform}
\newacronym{CTF}{CTF}{convolutive transfer function}
\newacronym{MTF}{MTF}{multiplicative transfer function}
\newacronym{RTF}{RTF}{relative transfer function}
\newacronym{MCLP}{MCLP}{multi channel linear prediction}
\newacronym{MPDR}{MPDR}{minimum power distortionless response}
\newacronym{MVDR}{MVDR}{minimum variance distortionless response}
\newacronym{LCMV}{LCMV}{linearly constrained minimum variance}
\newacronym{WPD}{WPD}{weighted power minimization distortionless response}
\newacronym{WPE}{WPE}{weighted prediction error}
\newacronym{TVG}{TVG}{time-varying complex circular Gaussian}
\newacronym{MISO}{MISO}{multiple-input single-output}
\newacronym{MIMO}{MIMO}{multiple-input multiple-output}
\newacronym{SPP}{SPP}{speech presence probability}
\newacronym{PESQ}{PESQ}{perceptual evaluation of speech quality}
\newacronym{FWSSNR}{FWSSNR}{frequency-weighted segmental signal-to-noise ratio}
\newacronym{CW}{CW}{covariance whitening}
\newacronym{VAD}{VAD}{voice activity detection}
\newacronym{SAD}{SAD}{source activity detection}
\newacronym{UCB}{UCB}{unified convolutional beamformer}
\newacronym{IRLS}{IRLS}{iteratively reweighted least squares}
\newacronym{MFMVDR}{MFMVDR}{multi-frame minimum variance distortionless response}
\newacronym{BMFMVDR}{BMFMVDR}{binaural MFMVDR}
\newacronym{TCN}{TCN}{temporal convolutional network}
\newacronym{DNN}{DNN}{deep neural network}
\newacronym{SNR}{SNR}{signal-to-noise ratio}
\newacronym{DRC}{DRC}{dynamic range compression}
\newacronym{HGR}{HGR}{half-gain rule}
\newacronym{CEC1}{CEC1}{Clarity Enhancement Challenge}
\newacronym{MBDRC}{MBDRC}{multi-band dynamic range compressor}
\newacronym{MBSTOI}{MBSTOI}{modified binaural short-term objective intelligibility}
\newacronym{SD-SDR}{SD-SDR}{scale-dependent signal-to-distortion ratio}
\newacronym{IFC}{IFC}{inter-frame correlation}
\newacronym{BTE}{BTE}{behind-the-ear}
\newacronym{SIR}{SIR}{signal-to-interferer ratio}
\newacronym{SINR}{SINR}{signal-to-interferer-and-noise ratio}
\newacronym{SRR}{SRR}{signal-to-reverberation ratio}
\newacronym{DUR}{DUR}{desired-to-undesired ratio}
\newacronym{wMPDR}{wMPDR}{weighted \gls{MPDR}}
\newacronym{SVD}{SVD}{singular value decomposition}
\newacronym{LS}{LS}{least-squares}
\newacronym{CBW}{CBW}{covariance blocking and whitening}
\newacronym{BOP}{BOP}{blind oblique projection}
\newacronym{CWu}{CWu}{\gls{CW} with the undesired covariance matrix}
\newacronym{PSD}{PSD}{power spectral density}

\DeclareMathOperator*{\argmin}{argmin}
\DeclarePairedDelimiter\abs{\lvert}{\rvert}%
\DeclarePairedDelimiter\norm{\lVert}{\rVert}%
\makeatletter
\let\oldabs\abs
\def\abs{\@ifstar{\oldabs}{\oldabs*}}
\let\oldnorm\norm
\def\norm{\@ifstar{\oldnorm}{\oldnorm*}}
\makeatother

\usepackage{siunitx}
\usepackage[hidelinks]{hyperref}

\title{Covariance Blocking and Whitening Method for Successive Relative Transfer Function Vector Estimation in Multi-Speaker Scenarios}

\name{Henri Gode and
      Simon Doclo\thanks{This work was funded by the Deutsche Forschungsgemeinschaft (DFG, German Research Foundation) -- Project ID 390895286 -- EXC 2177/1.}\vspace{-0.05cm}}
\address{{Dept. of Medical Physics and Acoustics and Cluster of Excellence Hearing4all,} \\
{University of Oldenburg, Germany, } 
 { \{henri.gode, simon.doclo\}@uni-oldenburg.de} \vspace{-0.05cm}}

\usepackage[subtle]{savetrees}

\begin{document}

\renewcommand{\sectionautorefname}{Section}
\renewcommand{\subsectionautorefname}{Section}
\renewcommand{\figureautorefname}{Figure}

\ninept
\maketitle

\begin{sloppy}

\begin{abstract}
  This paper addresses the challenge of estimating the \gls{RTF} vectors of multiple speakers in a noisy and reverberant environment. More specifically, we consider a scenario where two speakers activate successively. In this scenario, the \gls{RTF} vector of the first speaker can be estimated in a straightforward way and the main challenge lies in estimating the \gls{RTF} vector of the second speaker during segments where both speakers are simultaneously active. To estimate the \gls{RTF} vector of the second speaker the so-called \gls{BOP} method determines the oblique projection operator that optimally blocks the second speaker. Instead of blocking the second speaker, in this paper we propose a \gls{CBW} method, which first blocks the first speaker and applies whitening using the estimated noise covariance matrix and then estimates the \gls{RTF} vector of the second speaker based on a singular value decomposition. When using the estimated \gls{RTF} vectors of both speakers in a linearly constrained minimum variance beamformer, simulation results using real-world recordings for multiple speaker positions demonstrate that the proposed \gls{CBW} method outperforms the conventional \gls{BOP} and covariance whitening methods in terms of signal-to-interferer-and-noise ratio improvement.
\end{abstract}


\glsresetall
\vspace{-0.45cm}
\section{Introduction}
\label{sec:intro}
In many hands-free speech communication systems such as hearing aids, mobile phones and smart speakers, interfering sounds and ambient noise may degrade the speech quality and intelligibility of the recorded microphone signals~\cite{beutelmann_prediction_2006}.
When multiple microphones are available, beamforming is a widely used technique to enhance a target speaker and suppress interfering speakers and noise~\cite{veen_beamforming_1988,doclo_multichannel_2015,gannot_consolidated_2017}. One commonly used beamforming technique is the \gls{LCMV} beamformer \cite{veen_beamforming_1988, hadad_binaural_2016, goessling_blcmv, zhang2019distributed}, which requires estimates of the \gls{RTF} vectors of the target speaker and interfering speakers, as well as an estimate of the noise covariance matrix.

Over the last decades, several methods have been proposed to estimate the \gls{RTF} vector of a single speaker in a noisy environment, e.g., based on (weighted) least-squares~\cite{Gannot2001, cohen_relativ_2004}, using covariance subtraction and covariance whitening~\cite{warsitz2007blind, markovich_multichannel_2009, serizel_low-rank_2014, varzandeh2017iterative, markovich-golan_performance_2018}, using manifold learning~\cite{brendel2022manifold} or by jointly estimating the RTF vector and power spectral densities~\cite{tammen2019joint,li2023joint}.
The state-of-the-art \gls{CW} method estimates the \gls{RTF} vector of a single speaker by de-whitening the principal eigenvector of the whitened noisy covariance matrix, where an estimate of the noise covariance matrix is used for the whitening.

In contrast to estimating the \gls{RTF} vector of a single speaker, estimating the \gls{RTF} vectors of multiple speakers that are simultaneously active is more challenging. In a multi-speaker scenario, the \gls{CW} method can only estimate a subspace spanning the \gls{RTF} vectors of all speakers, instead of the individual \gls{RTF} vectors~\cite{markovich_multichannel_2009}. Methods to estimate the \gls{RTF} vectors of multiple speakers have been proposed, e.g., using an expectation maximization algorithm~\cite{schwartz2017two}, utilizing subframes to perform factor analysis~\cite{koutrouvelis2019robust} or using joint diagonalization~\cite{Changheng_LCAMSRTFE_2022}. A drawback of the methods is the permutation ambiguity, requiring a method to correctly assign the estimated \gls{RTF} vectors to the speakers in each frequency band. An \gls{RTF} vector update method
using Procrustes analysis was proposed in~\cite{dietzen2020square}.

\begin{figure}[t]
  \centering
  \centerline{\includegraphics[width=\columnwidth]{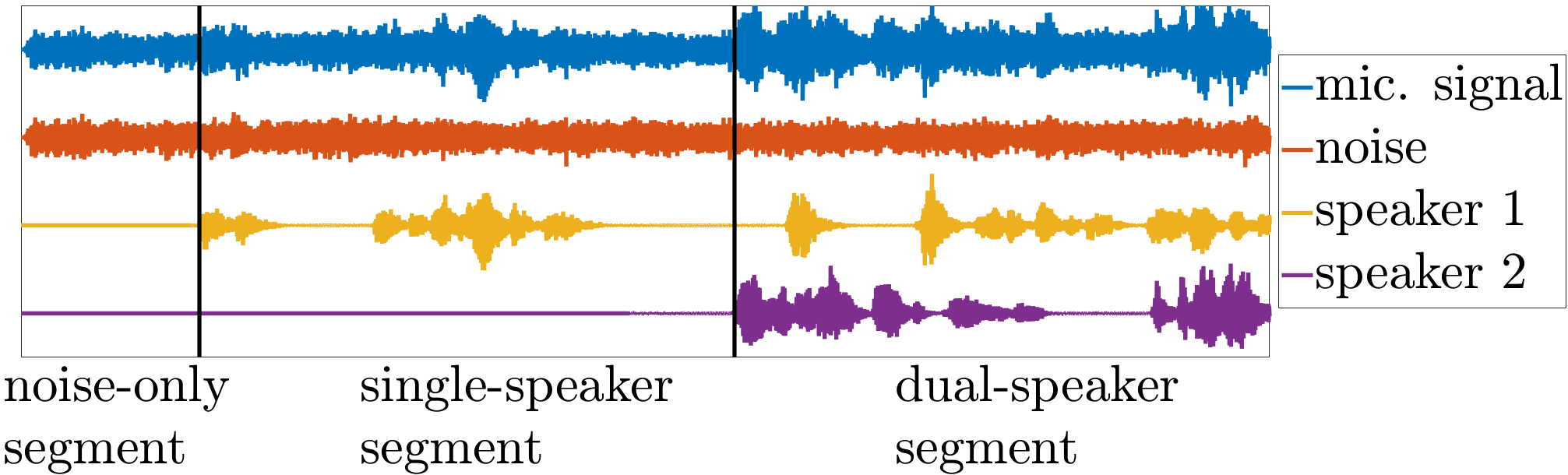}}
  \caption{Example of successive speaker scenario with two speakers, depicting a microphone signal and its speech and noise components for three different segments.}
  \label{fig:segments}
  \vspace{-0.2cm}
\end{figure}

In this paper, we consider a specific scenario with two speakers that activate successively (see \autoref{fig:segments}).
In this scenario, the \gls{RTF} vector of the first speaker can be estimated in the single-speaker segment, e.g., using the \gls{CW} method with an estimate of the noise covariance matrix from the noise-only segment. The focus of this paper is on estimating the \gls{RTF} vector of the second speaker in the dual-speaker segment, using available estimates of the noise covariance matrix and the \gls{RTF} vector of the first speaker. First, we consider a modified version of the conventional \gls{CW} method, where the whitening is performed using the undesired covariance matrix (estimated in the single-speaker segment) instead of the noise covariance matrix. A disadvantage of this method is its dependence on the possibly time-varying \gls{PSD} of the first speaker. Second, we consider the \gls{BOP} method~\cite{cherkassky_successive_2020}, which estimates the \gls{RTF} vector of the second speaker by determining the oblique projection operator which optimally blocks the unknown second speaker. 
A disadvantage of the \gls{BOP} method is that it assumes a sufficiently large \gls{SNR}.
Third, we propose the \gls{CBW} method, which overcomes the mentioned disadvantages of the \gls{CW} and \gls{BOP} methods, i.e. it is independent of the \gls{PSD} of the first speaker and does not assume a large \gls{SNR}. 
Instead of blocking the unknown second speaker as done by the \gls{BOP} method, the \gls{CBW} method uses a residual maker matrix to block the known \gls{RTF} vector of the first speaker, followed by noise whitening. Since the blocking of the first speaker leads to a dimension reduction, it is shown that a \gls{SVD} instead of an eigenvalue decomposition is required to extract the \gls{RTF} vector of the second speaker.
When using the estimated \gls{RTF} vectors in an \gls{LCMV} beamformer, simulation results using real-world recordings for multiple speaker positions and \glspl{SIR} demonstrate that the proposed \gls{CBW} method outperforms  the conventional methods in terms of \gls{SINR} improvement.

\section{Signal Model}
\label{sec:signal_model}
We consider a scenario where two speakers are successively activated and captured by an array with $M$ microphones in a noisy and reverberant environment. \autoref{fig:segments} depicts this specific scenario, which can be divided into three time segments: a noise-only segment $(s=1)$, a single-speaker segment containing the first speaker and noise $(s=2)$, and a dual-speaker segment containing both speakers and noise $(s=3)$, where $s$ denotes the segment index. We assume that the segment boundaries are known. Without loss of generality, the first speaker is considered the interfering speaker, whereas the second speaker is considered the target speaker. We assume that for the whole signal duration the scenario is spatially stationary, i.e., the speakers do not move.
The \gls{STFT} coefficients of the microphone signals at time frame $t$ are denoted by
\begin{align}
    \mathbf{y}_t = \begin{bmatrix} y_{1,t} & \cdots & y_{M,t} \end{bmatrix}^\mathrm{T} \in \mathbb{C}^{M\times 1},
    \label{eq:sig_vec_def}
\end{align}
where $\{\cdot\}^{\mathrm{T}}$ denotes the transpose operator. The frequency index is omitted as it is assumed that each frequency band is independent and can be processed individually.
The multi-channel microphone signal $\mathbf{y}_t$ can be written as the sum of the target speaker component $\mathbf{x}_t$ and the undesired component $\mathbf{v}_t$, which consists of the interfering speaker component $\mathbf{u}_t$ and the noise component $\mathbf{n}_t$, i.e.,
\begin{align}
\mathbf{y}_t = \mathbf{x}_t + \underbrace{\mathbf{u}_t + \mathbf{n}_t}_{\mathbf{v}_t} = \mathbf{h}x_{r,t} + \mathbf{g}u_{r,t} + \mathbf{n}_t,
\label{eq:direct_sig_mod}
\end{align}
where the vectors $\mathbf{x}_t$, $\mathbf{u}_t$, $\mathbf{n}_t$ and $\mathbf{v}_t$ are defined similarly as in~\eqref{eq:sig_vec_def}.
Assuming sufficiently large \gls{STFT} frames, the speaker components $\mathbf{x}_t$ and $\mathbf{u}_t$ are modeled as the multiplication of the STFT coefficient of a reference microphone (denoted by $r$) 
with the respective time-invariant \gls{RTF} vectors $\mathbf{h}$ and $\mathbf{g}$~\cite{avargel_multiplicative_2007}. Note that the reference entry of both \gls{RTF} vectors equals $1$.
Assuming all signal components in~\eqref{eq:direct_sig_mod} to be uncorrelated, the noisy covariance matrix for each time segment $s$ is given by
\begin{align}
\mathbf{R}_{\mathbf{y},s}
= \underbrace{\mathbf{h}\phi_{x,s}\mathbf{h}^{\mathrm{H}}}_{\mathbf{R}_{\mathbf{x},s}}+ \underbrace{\underbrace{\mathbf{g}\phi_{u, s}\mathbf{g}^{\mathrm{H}}}_{\mathbf{R}_{\mathbf{u}, s}} + \mathbf{R}_{\mathbf{n}}}_{\mathbf{R}_{\mathbf{v}, s}},
\label{eq:sig_model}
\end{align}
where $\{\cdot\}^{\mathrm{H}}$ denotes the conjugate transpose operator. The matrices $\mathbf{R}_{\mathbf{u}, s}$ and $\mathbf{R}_{\mathbf{x}, s}$ denote the rank-1 covariance matrices of the first and second speaker, respectively, $\mathbf{R}_{\mathbf{n}}$ and $\mathbf{R}_{\mathbf{v},s}$ denote the full-rank covariance matrices of the noise and undesired component, respectively, and $\phi_{x,s}=\mathcal{E}\{\abs{x_r}^{2}\}$ and $\phi_{u,s}=\mathcal{E}\{\abs{u_r}^{2}\}$ denote the \glspl{PSD} with $\mathcal{E}\{\cdot\}$ the expectation operator. Note that $\phi_{x,1} = \phi_{u,1} = \phi_{x,2} = 0$ for the considered successive speaker scenario.

\section{LCMV beamformer}
\label{sec:LCMV}
To extract the second speaker (target speaker) and suppress the first speaker (interfering speaker) and noise in the dual-speaker segment $(s=3)$, we will use an \gls{LCMV} beamformer.
The \gls{LCMV} beamformer minimizes the noise \gls{PSD} subject to linear constraints, which aim at keeping the target speaker distortionless and reducing the interfering speaker~\cite{veen_beamforming_1988, hadad_binaural_2016, goessling_blcmv}, i.e.,
\begin{align}
    \mathbf{w} = \argmin_{\tilde{\mathbf{w}}} \left(\tilde{\mathbf{w}}^{\mathrm{H}}\mathbf{R}_{\mathbf{n}}\tilde{\mathbf{w}}\right)
    \quad\mathrm{s.t.}\quad
    \begin{matrix}
        \tilde{\mathbf{w}}^{\mathrm{H}}\mathbf{x}_t = x_{r,t} \\
        \tilde{\mathbf{w}}^{\mathrm{H}}\mathbf{u}_t = \delta u_{r,t}
    \end{matrix},
    \label{eq:beamforming_opt_problem}
\end{align}
where $\delta$ denotes the scaling factor to reduce the interfering speaker.
The widely known solution to the minimization problem in~\eqref{eq:beamforming_opt_problem} is given by
\begin{align}
    \boxed{\mathbf{w} = \mathbf{R}_{\mathbf{n}}^{-1}\mathbf{C}\left(\mathbf{C}^{\mathrm{H}}\mathbf{R}_{\mathbf{n}}^{-1}\mathbf{C}\right)^{-1}\begin{bmatrix}
        1 \\ \delta
    \end{bmatrix}}
    \label{eq:LCMV_beamformer}
\end{align}
where the constraint matrix $\mathbf{C}$ contains the \gls{RTF} vectors of both speakers, i.e., $\mathbf{C} = \begin{bmatrix}   \mathbf{h} & \mathbf{g}\end{bmatrix}$.
Applying the \gls{LCMV} beamformer to the microphone signals yields the enhanced signal
\begin{align}
    \mathbf{z}_{t} = \mathbf{w}^{\mathrm{H}}\mathbf{y}_{t}.
    \label{eq:LCMV_beamforming}
\end{align}
As can be seen in~\eqref{eq:LCMV_beamformer}, the \gls{LCMV} beamformer requires an estimate of the noise covariance matrix $\mathbf{R}_{\mathbf{n}}$, which can be easily obtained in the noise-only segment, and estimates of the \gls{RTF} vectors of both speakers. \autoref{sec:CW_interfering} discusses a method to estimate the \gls{RTF} vector $\mathbf{g}$ of the first speaker. The main focus of the paper is on estimating the \gls{RTF} vector $\mathbf{h}$ of the second speaker, for which several methods will be presented in~\autoref{sec:conventional_methods} and~\autoref{sec:proposed_method}.

\subsection{Covariance Whitening (CW)}
\label{sec:CW_interfering}

In the single-speaker segment $(s=2)$ where $\phi_{x,2}=0$, the \gls{RTF} vector $\mathbf{g}$ of the first speaker can be estimated using the state-of-the-art \gls{CW} method~\cite{markovich_multichannel_2009, serizel_low-rank_2014, markovich-golan_performance_2018}. First, the noisy covariance matrix $\mathbf{R}_{\mathbf{y},2}$ in~\eqref{eq:sig_model} is whitened using a square-root decomposition of the noise covariance matrix $\mathbf{R}_{\mathbf{n}} = \mathbf{R}_{\mathbf{n}}^{\nicefrac{\mathrm{H}}{2}}\mathbf{R}_{\mathbf{n}}^{\nicefrac{1}{2}}$, i.e.,
\begin{align}
    \mathbf{R}_{\mathbf{n}}^{-\nicefrac{\mathrm{H}}{2}}\mathbf{R}_{\mathbf{y},2}\mathbf{R}_{\mathbf{n}}^{-\nicefrac{1}{2}} = \mathbf{R}_{\mathbf{n}}^{-\nicefrac{\mathrm{H}}{2}}\mathbf{g}\phi_{u,2}\mathbf{g}^{\mathrm{H}}\mathbf{R}_{\mathbf{n}}^{-\nicefrac{1}{2}} + \mathbf{I}_{M}.
    \label{eq:CW_noise_whitening}
\end{align}
This whitening operation spatially decorrelates the noise, transforming the noise covariance matrix into an identity matrix. From~\eqref{eq:CW_noise_whitening}, the \gls{RTF} vector $\mathbf{g}$ can then be estimated as the normalized de-whitened principal eigenvector of the whitened noisy covariance matrix, i.e.,
\begin{align}
    \boxed{\widehat{\mathbf{g}} = \nicefrac{\tilde{\mathbf{g}}}{\mathbf{e}_r^{\mathrm{T}}\tilde{\mathbf{g}}}
    \quad\mathrm{with}\quad
    \tilde{\mathbf{g}} = \mathbf{R}_{\mathbf{n}}^{\nicefrac{\mathrm{H}}{2}}\mathcal{P}\{\mathbf{R}_{\mathbf{n}}^{-\nicefrac{\mathrm{H}}{2}}\mathbf{R}_{\mathbf{y}, 2}\mathbf{R}_{\mathbf{n}}^{-\nicefrac{1}{2}}
    \}}
    \label{eq:CW_interfering}
\end{align}
where $\mathbf{e}_{r} = \begin{bmatrix}
    0 & \cdots & 1 & \cdots & 0
\end{bmatrix}^{\mathrm{T}}
$ is an $M$-dimensional selection vector of the reference microphone $r$ and $\mathcal{P}\{\cdot\}$ denotes the principal eigenvector operator.



\section{Conventional RTF Estimation Methods}
\label{sec:conventional_methods}
In this section, we describe two conventional methods to estimate the \gls{RTF} vector $\mathbf{h}$ of the second speaker in the dual-speaker segment, where both speakers and noise are present, assuming estimates of the noise covariance matrix $\mathbf{R}_{\mathbf{n}}$, the undesired covariance matrix $\mathbf{R}_{\mathbf{v}}$ and the \gls{RTF} vector $\mathbf{g}$ of the first speaker to be available.

\subsection{CW with undesired covariance matrix (CWu)}
\label{sec:CW}
In \autoref{sec:CW_interfering} the \gls{CW} method was discussed to estimate the \gls{RTF} vector of the first speaker. Although the \gls{CW} method typically performs whitening using the estimated noise covariance matrix $\mathbf{R}_{\mathbf{n}}$, this method can also be used to estimate the \gls{RTF} vector of the second speaker in the dual-speaker segment by performing whitening with the undesired covariance matrix $\mathbf{R}_{\mathbf{v}}$ (estimated during the single-speaker segment), i.e.,
\begin{align}
    \boxed{\widehat{\mathbf{h}}^{(\mathrm{CW})} = \nicefrac{\tilde{\mathbf{h}}}{\mathbf{e}_r^{\mathrm{T}}\tilde{\mathbf{h}}}
    \quad\mathrm{with}\quad
    \tilde{\mathbf{h}} = \mathbf{R}_{\mathbf{v},2}^{\nicefrac{\mathrm{H}}{2}}\mathcal{P}\{\mathbf{R}_{\mathbf{v},2}^{-\nicefrac{\mathrm{H}}{2}}\mathbf{R}_{\mathbf{y},3}\mathbf{R}_{\mathbf{v},2}^{-\nicefrac{1}{2}}
    \}}
    \label{eq:CW}
\end{align}
This will be referred to as \gls{CWu}.
It should be noted that since the \gls{PSD} of the first speaker in the single-speaker segment $\phi_{u,2}$ may not be the same as in the dual-speaker segment $\phi_{u,3}$, the undesired covariance matrix $\mathbf{R}_{\mathbf{v}, 2}$ used for whitening may strongly deviate form the undesired covariance matrix $\mathbf{R}_{\mathbf{v}, 3}$, resulting in a biased \gls{RTF} vector estimate.

\subsection{Blind Oblique Projection (BOP)}
\label{sec:BOP}
Avoiding the influence of the time-varying \gls{PSD} $\phi_u$ of the first speaker, the \gls{BOP} method proposed in~\cite{cherkassky_successive_2020} estimates the \gls{RTF} vector $\mathbf{h}$ of the second speaker by blocking the second speaker as much as possible while keeping the first speaker distortionless. 
This can be achieved by the so-called oblique projection operator~\cite{behrens_signal_1994} with vector variable $\bm{\theta}$, defined as
\begin{align}
    \mathbf{P}^{\angle}_{\mathbf{g}\bm{\theta}}
    = \mathbf{g}\left(\mathbf{g}^{\mathrm{H}}\mathbf{P}^{\perp}_{\bm{\theta}}\mathbf{g}\right)^{-1}\mathbf{g}^{\mathrm{H}}\mathbf{P}^{\perp}_{\bm{\theta}},
    \label{eq:oblique_proj_op}
\end{align}
where $\mathbf{P}^{\perp}_{\bm{\theta}}$ denotes the residual maker matrix 
\begin{align}
    \mathbf{P}^{\perp}_{\bm{\theta}} = \mathbf{I}_{M} - \nicefrac{\bm{\theta}\bm{\theta}^{\mathrm{H}}}{\bm{\theta}^{\mathrm{H}}\bm{\theta}}.
\end{align}
The oblique projection operator in~\eqref{eq:oblique_proj_op} keeps the \gls{RTF} vector of the first speaker, i.e., $\mathbf{P}^{\angle}_{\mathbf{g}\bm{\theta}}\mathbf{g} = \mathbf{g}$, but blocks the range of the vector $\bm{\theta}$, i.e., $\mathbf{P}^{\angle}_{\mathbf{g}\bm{\theta}}\bm{\theta} = \mathbf{0}_{M\times 1}$. Applying the oblique projection operator in~\eqref{eq:oblique_proj_op} to \eqref{eq:sig_model} in the dual-speaker segment and taking the trace yields
\begin{multline}
    \mathrm{Tr}\{\mathbf{P}^{\angle}_{\mathbf{g}\bm{\theta}}\mathbf{R}_{\mathbf{y},3}\mathbf{P}^{\angle\mathrm{H}}_{\mathbf{g}\bm{\theta}}\}
    = \phi_{x,3}\mathbf{h}^{\mathrm{H}}\mathbf{P}^{\angle\mathrm{H}}_{\mathbf{g}\bm{\theta}}\mathbf{P}^{\angle}_{\mathbf{g}\bm{\theta}}\mathbf{h} + \phi_{u,3}\mathbf{g}^{\mathrm{H}}\mathbf{g}
    \\
    + \mathrm{Tr}\{\mathbf{P}^{\angle}_{\mathbf{g}\bm{\theta}}\mathbf{R}_{\mathbf{n}}\mathbf{P}^{\angle\mathrm{H}}_{\mathbf{g}\bm{\theta}}\}.
    \label{eq:BOP_sig_mod}
\end{multline}
In~\cite{cherkassky_successive_2020} a sufficiently large \gls{SNR} was assumed, such that the noise term $\mathrm{Tr}\{\mathbf{P}^{\angle}_{\mathbf{g}\bm{\theta}}\mathbf{R}_{\mathbf{n}}\mathbf{P}^{\angle\mathrm{H}}_{\mathbf{g}\bm{\theta}}\}$ in~\eqref{eq:BOP_sig_mod} can be neglected.
Since the term $\phi_{u,3}\mathbf{g}^{\mathrm{H}}\mathbf{g}$ is a constant, it can be seen that the power of the projected noisy covariance matrix $\mathrm{Tr}\{\mathbf{P}^{\angle}_{\mathbf{g}\bm{\theta}}\mathbf{R}_{\mathbf{y},3}\mathbf{P}^{\angle\mathrm{H}}_{\mathbf{g}\bm{\theta}}\}$ is minimized when the oblique projection operator $\mathbf{P}^{\angle}_{\mathbf{g}\bm{\theta}}$ blocks the \gls{RTF} vector $\mathbf{h}$ of the second speaker in~\eqref{eq:BOP_sig_mod}, which is achieved if the vector $\bm{\theta}$ points in the same direction as $\mathbf{h}$.
Therefore an estimate of the \gls{RTF} vector of the second speaker can be obtained as
\begin{align}
    \boxed{\widehat{\mathbf{h}}^{(\mathrm{BOP})} = \nicefrac{\tilde{\mathbf{h}}}{\mathbf{e}_r^{\mathrm{T}}\tilde{\mathbf{h}}}
    \quad\mathrm{with}\quad
    \tilde{\mathbf{h}} = \argmin_{\bm{\theta}}\left(\mathrm{Tr}\{\mathbf{P}^{\angle}_{\mathbf{g}\bm{\theta}}\mathbf{R}_{\mathbf{y},3}\mathbf{P}^{\angle\mathrm{H}}_{\mathbf{g}\bm{\theta}}\}\right)}
    \label{eq:BOP_min_problem}
\end{align}
Since no closed-form solution for the optimization problem in~\eqref{eq:BOP_min_problem} exists, it was proposed in~\cite{cherkassky_successive_2020} to use a gradient-descent method. For faster computational speed and higher robustness, in this paper we provided the analytical gradient of~\eqref{eq:BOP_min_problem} to the sequential-quadratic-programming method~\cite{nocedal1999numerical}, implemented within the \textsc{Matlab} optimization toolbox~\cite{MatlabOptimization}.

\section{Covariance Blocking and Whitening}
\label{sec:proposed_method}

Instead of whitening using the undesired covariance matrix $\mathbf{R}_{\mathbf{v}}$ as in the \gls{CWu} method, in this section we propose a method that blocks the known first speaker before noise whitening. This can be seen as introducing blocking of the first speaker to the conventional \gls{CW} method, to enable estimation of the individual \gls{RTF} vector of the second speaker. It should be noted that this blocking approach differs fundamentally from the \gls{BOP} method, where the unknown second speaker is blocked instead of the known first speaker. In principle, the \gls{RTF} vector $\mathbf{g}$ of the first speaker can be blocked by applying the residual maker matrix
\begin{align}
    \mathbf{P}^{\perp}_{{ \mathbf{g}}} = \mathbf{I}_{M} - \nicefrac{\mathbf{g}\mathbf{g}^{\mathrm{H}}}{\left(\mathbf{g}^{\mathrm{H}}\mathbf{g}\right)}
    \label{eq:res_make_mat}
\end{align}
with $\mathbf{g}^{\mathrm{H}}\mathbf{P}^{\perp}_{{ \mathbf{g}}} = \mathbf{0}_{1\times M}$, to~\eqref{eq:sig_model} from the right in the dual-speaker segment, yielding
\begin{align}
    \mathbf{R}_{\mathbf{y},3} \mathbf{P}^{\perp}_{\mathbf{g}}=  {\mathbf{h}}\phi_{x,3}{\mathbf{h}}^{\mathrm{H}}\mathbf{P}^{\perp}_{{ \mathbf{g}}} + \mathbf{R}_{\mathbf{n}}\mathbf{P}^{\perp}_{{ \mathbf{g}}}.
    \label{eq:blocked_sig_mod_not_reduced}
\end{align}
It should be noted that since the matrix $\mathbf{P}^{\perp}_{{ \mathbf{g}}}$ in~\eqref{eq:blocked_sig_mod_not_reduced} has rank $(M-1)$, the noise term $\mathbf{R}_{\mathbf{n}}\mathbf{P}^{\perp}_{{ \mathbf{g}}}$ in~\eqref{eq:blocked_sig_mod_not_reduced} also has rank $(M-1)$. Since full column rank of the noise term is required to whiten the noise component, we propose to use a dimension-reduced version $\mathbf{P}^{\perp}_{{\mathbf{g},\mathrm{r}}} = \mathbf{P}^{\perp}_{{\mathbf{g}}}\begin{bmatrix}
\mathbf{I}_{M-1} & \mathbf{0}_{\left(M-1\right)\times 1}\end{bmatrix}^{\mathrm{T}} \in \mathbb{C}^{M\times M-1}$ of the residual maker matrix $\mathbf{P}^{\perp}_{{ \mathbf{g}}}$ instead, i.e.,
\begin{align}
    \mathbf{R}_{\mathbf{y},3} \mathbf{P}^{\perp}_{\mathbf{g},\mathrm{r}}=  {\mathbf{h}}\phi_{x,3}{\mathbf{h}}^{\mathrm{H}}\mathbf{P}^{\perp}_{{ \mathbf{g},\mathrm{r}}} + \mathbf{R}_{\mathbf{n}}\mathbf{P}^{\perp}_{{ \mathbf{g},\mathrm{r}}} \in \mathbb{C}^{M\times M-1}.
    \label{eq:blocked_sig_mod}
\end{align}
Whitening the blocked noise term $\mathbf{R}_{\mathbf{n}}\mathbf{P}^{\perp}_{{ \mathbf{g},\mathrm{r}}}$ in~\eqref{eq:blocked_sig_mod} using its pseudo-inverse denoted by $\{\cdot\}^{+}$, i.e., transforming it to an identity matrix similarly as for the conventional \gls{CW} method in~\eqref{eq:CW_noise_whitening}, and subtracting the resulting identity matrix yields
\begin{align}
    \mathbf{R}_{\mathbf{y}}^{\mathrm{w}} = \left(\mathbf{R}_{\mathbf{n}}\mathbf{P}^{\perp}_{{\mathbf{g},\mathrm{r}}}\right)^{+}\mathbf{R}_{\mathbf{y},3} \mathbf{P}^{\perp}_{{\mathbf{g},\mathrm{r}}} - \mathbf{I}_{M-1}
     =  \left(\mathbf{R}_{\mathbf{n}}\mathbf{P}^{\perp}_{{\mathbf{g},\mathrm{r}}}\right)^{+}{\mathbf{h}}\phi_{x}{\mathbf{h}}^{\mathrm{H}}\mathbf{P}^{\perp}_{{\mathbf{g},\mathrm{r}}}.
    \label{eq:CBW_sig_mod_reformulated}
\end{align}
This set of equations has dimensions $(M-1)\times (M-1)$ in contrast to the signal model in~\eqref{eq:sig_model} having dimensions $M\times M$. It should be noted that the right side of~\eqref{eq:CBW_sig_mod_reformulated} is given by the outer product of two transformed versions of the \gls{RTF} vector $\mathbf{h}$ of the second speaker, whereby both transformations only depend on $\mathbf{R}_{\mathbf{n}}$ and $\mathbf{g}$.
The transformed versions of the \gls{RTF} vector can be extracted by means of an \gls{SVD}, i.e.,
\begin{align}
    \mathbf{q}_{\mathrm{L}} =& \;\mathcal{S}_{\mathrm{L}}\{ \mathbf{R}_{\mathbf{y}}^{\mathrm{w}}\} = \left(\mathbf{R}_{\mathbf{n}}\mathbf{P}^{\perp}_{{\mathbf{g},\mathrm{r}}}\right)^{+}{\mathbf{h}}\alpha_{\mathrm{L}},
    \label{eq:q_L}
    \\
    \mathbf{q}_{\mathrm{R}} =& \;\mathcal{S}_{\mathrm{R}}\{\mathbf{R}_{\mathbf{y}}^{\mathrm{w}}\} = \left(\mathbf{P}^{\perp}_{{\mathbf{g},\mathrm{r}}}\right)^{\mathrm{H}}{\mathbf{h}}\alpha_{\mathrm{R}},
    \label{eq:q_R}
\end{align}
where $\mathcal{S}_{\mathrm{L}}\{\cdot\}$ and $\mathcal{S}_{\mathrm{R}}\{\cdot\}$ denote the principal left and right singular vector operator, respectively, and $\alpha_{\mathrm{L}}$ and $\alpha_{\mathrm{R}}$ are scaling factors.
It should be noted that since $\mathbf{q}_{\mathrm{L}}$ and $\mathbf{q}_{\mathrm{R}}$ are $(M-1)$-dimensional vectors, neither vector provides a solution for the $M$-dimensional \gls{RTF} vector $\mathbf{h}$.

By introducing the scaled \gls{RTF} vector $\tilde{\mathbf{h}} = \alpha_{\mathrm{L}}\mathbf{h}$ and the weighting factor $\alpha=\nicefrac{\alpha_\mathrm{L}}{\alpha_\mathrm{R}}$ and by stacking~\eqref{eq:q_L} and~\eqref{eq:q_R} into one unified set of equations, we obtain
\begin{align}
    \begin{bmatrix}
    \mathbf{q}_{\mathrm{L}} \\
    \mathbf{q}_{\mathrm{R}}\alpha
    \end{bmatrix} \overset{!}{=}  \mathbf{B}
    \tilde{\mathbf{h}}
    \quad\mathrm{with}\quad \mathbf{B}= \begin{bmatrix}
    \left(\mathbf{R}_{\mathbf{n}}\mathbf{P}_{\mathbf{g}, \mathrm{r}}^{\perp}\right)^{+} \\
    \left(\mathbf{P}_{\mathbf{g}, \mathrm{r}}^{\perp}\right)^{\mathrm{H}}
    \end{bmatrix}.
    \label{eq:non_lin_eq_sys}
\end{align}
This non-linear set of equations contains $2(M-1)$ equations and $M+1$ unknowns ($\tilde{\mathbf{h}}$ and $\alpha$), hence requiring $2(M-1) \geq M+1$, i.e., $M\geq 3$.
Inverting $\mathbf{B}$ using its pseudo-inverse and applying it to~\eqref{eq:non_lin_eq_sys} from the left yields
\begin{align}
    \tilde{\mathbf{h}}\overset{}{=}  \mathbf{B}^{+}\begin{bmatrix}
    \mathbf{q}_{\mathrm{L}} \\
    \mathbf{q}_{\mathrm{R}}\alpha
    \end{bmatrix},
    \label{eq:h_solution}
\end{align}
which depends on the unknown $\alpha$.
Substituting~\eqref{eq:h_solution} into~\eqref{eq:non_lin_eq_sys} leads to
\begin{align}
    \begin{bmatrix}
    \mathbf{q}_{\mathrm{L}} \\
    \mathbf{q}_{\mathrm{R}}\alpha
    \end{bmatrix} \overset{!}{=}  \mathbf{B}
    \mathbf{B}^{+}\begin{bmatrix}
    \mathbf{q}_{\mathrm{L}} \\
    \mathbf{q}_{\mathrm{R}}\alpha
    \end{bmatrix}
    \quad\Rightarrow\quad
    \mathbf{P}^{\perp}_{\mathbf{B}}\begin{bmatrix}
    \mathbf{q}_{\mathrm{L}} \\
    \mathbf{q}_{\mathrm{R}}\alpha
    \end{bmatrix} \overset{!}{=} \mathbf{0}_{2(M-1) \times 1},
    \label{eq:alpha_eq_sys}
\end{align}
where $\mathbf{P}^{\perp}_{\mathbf{B}}$ is the residual maker matrix of $\mathbf{B}$ defined similarly to~\eqref{eq:res_make_mat}. Reformulating~\eqref{eq:alpha_eq_sys} using the block decomposition $\mathbf{P}^{\perp}_{\mathbf{B}} = \begin{bmatrix}
    \mathbf{P}^{\perp,\mathrm{L}}_{\mathbf{B}} & \mathbf{P}^{\perp,\mathrm{R}}_{\mathbf{B}}
\end{bmatrix}$ provides a solution for $\alpha$, i.e.,
\begin{align}
    -\mathbf{P}^{\perp,\mathrm{L}}_{\mathbf{B}}\mathbf{q}_{\mathrm{L}} = \mathbf{P}^{\perp,\mathrm{R}}_{\mathbf{B}}\mathbf{q}_{\mathrm{R}}\alpha
    \quad\Rightarrow\quad
    \alpha = -\left(\mathbf{P}^{\perp,\mathrm{R}}_{\mathbf{B}}\mathbf{q}_{\mathrm{R}}\right)^{+}\mathbf{P}^{\perp,\mathrm{L}}_{\mathbf{B}}\mathbf{q}_{\mathrm{L}}.
    \label{eq:alpha_solution}
\end{align}
Substituting $\alpha$ in~\eqref{eq:alpha_solution} into $\tilde{\mathbf{h}}$ in~\eqref{eq:h_solution} and applying normalization provides an estimate of the \gls{RTF} vector of the second speaker, i.e.,
\begin{align}
    \boxed{\widehat{\mathbf{h}}^{(\mathrm{CBW})} = \nicefrac{\tilde{\mathbf{h}}}{\mathbf{e}_{r}^{\mathrm{T}}\tilde{\mathbf{h}}}
    \quad\mathrm{with}\quad
    \tilde{\mathbf{h}}\overset{}{=}  \mathbf{B}^{+}\begin{bmatrix}
    \mathbf{q}_{\mathrm{L}} \\
    -\mathbf{q}_{\mathrm{R}}\left(\mathbf{P}^{\perp,\mathrm{R}}_{\mathbf{B}}\mathbf{q}_{\mathrm{R}}\right)^{+}\mathbf{P}^{\perp,\mathrm{L}}_{\mathbf{B}}\mathbf{q}_{\mathrm{L}}
    \end{bmatrix}}
    \label{eq:LS_solution}
\end{align}

\section{Experimental Results}
\label{sec:evaluation}
In this section, we compare the performance of the conventional \gls{RTF} vector estimation methods (see~\autoref{sec:conventional_methods}) with the proposed \gls{CBW} method, when using the estimated \gls{RTF} vectors of the target and interfering speakers (second and first speaker, respectively) in an \gls{LCMV} beamformer.

\subsection{Acoustic Scenario}

\autoref{fig:scenario} depicts the acoustic setup. We considered a linear array with 4 microphones with a microphone spacing of \SI{2}{\cm}, located approximately in the center of an acoustic laboratory $(\SI{7}{\m}\times \SI{6}{\m} \times \SI{2.7}{\m})$ with a reverberation time $T_{60}\approx \SI{500}{\milli\s}$. The acoustic scenario consists of constantly active background noise, one interfering speaker active in the interval $\left[\SI{1}{\s}, \SI{7}{\s}\right]$ and one target speaker active in the interval $\left[\SI{4}{\s}, \SI{7}{\s}\right]$. The target and interfering speech components at the microphones were generated by convolving clean speech signals with room impulse responses measured from loudspeakers at 9 different positions (see~\autoref{fig:scenario}). Quasi-diffuse noise was generated by playing back uncorrelated babble noise using 4 loudspeakers facing the corners of the laboratory.
The sampling frequency was equal to \SI{16}{\kilo\hertz}.

\begin{figure}[t]
  \centering
  \centerline{\includegraphics[width=\columnwidth]{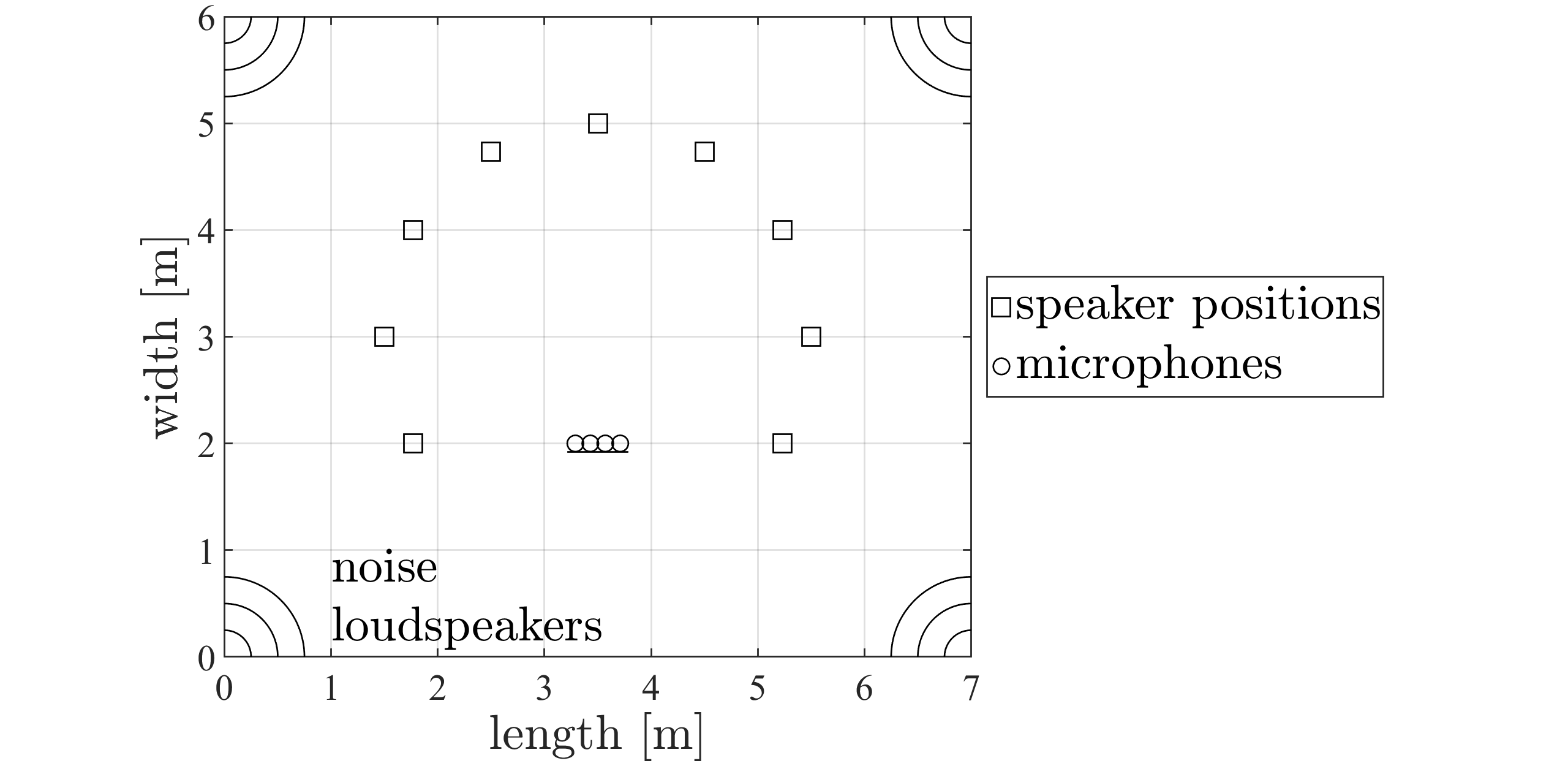}}
  \caption{Acoustic setup with 9 speaker positions, a linear microphone array with 4 microphones and background noise.}
  \label{fig:scenario}
\end{figure}

The \gls{LCMV} beamformer and \gls{RTF} estimation is performed within an \gls{STFT} framework with a frame length of 3200 samples (corresponding to $\SI{200}{\milli\s}$), a frame shift of 800 samples (corresponding to $\SI{50}{\milli\s}$) and a square-root-Hann window for analysis and synthesis.
The scaling factor of the interfering speaker in \eqref{eq:LCMV_beamformer} was set to $\delta = \SI{-40}{\dB}$. The noise covariance matrix $\mathbf{R}_{\mathbf{n}}$ and the undesired covariance matrix $\mathbf{R}_{\mathbf{v},2}$ were estimated using the sample covariance matrix method in the noise-only and the single-speaker segment, respectively. The \gls{RTF} vector $\mathbf{g}$ of the interfering speaker was estimated by \gls{CW} in the single-speaker segment as described in~\autoref{sec:CW_interfering}.

The performance of the \gls{LCMV} beamformer is evaluated in terms of the broadband \glsfirst{SINR} improvement for each microphone as reference ($r\in\{1, \ldots,M\}$), i.e.,
\begin{align}
    \Delta \mathrm{SINR}_r = 10 \cdot\log_{10}\left( \frac{\sum_{n} \abs{x_{r,n}^{\mathrm{out}}}^2}{\sum_{n} \abs{v_{r,n}^{\mathrm{out}}}^2}\right) -10 \cdot\log_{10}\left( \frac{\sum_{n} \abs{x_{r,n}^{\mathrm{in}}}^2}{\sum_{n} \abs{v_{r,n}^{\mathrm{in}}}^2}\right),
\end{align}
where $n$ denotes the sample index, $x_{r,n}^{\mathrm{in}}$ and $v_{r,n}^{\mathrm{in}}$ denote the time-domain target speaker component and the undesired component (sum of interfering speaker and noise) in the $r$-th microphone signal, respectively, and $x_{r,n}^{\mathrm{out}}$ and $v_{r,n}^{\mathrm{out}}$ denote the shadow-filtered versions of $x_{r,n}^{\mathrm{in}}$ and $v_{r,n}^{\mathrm{in}}$ using the \gls{LCMV} beamformer in~\eqref{eq:LCMV_beamforming}.

\begin{figure}[t]
  \centering
  \centerline{\includegraphics[width=\columnwidth]{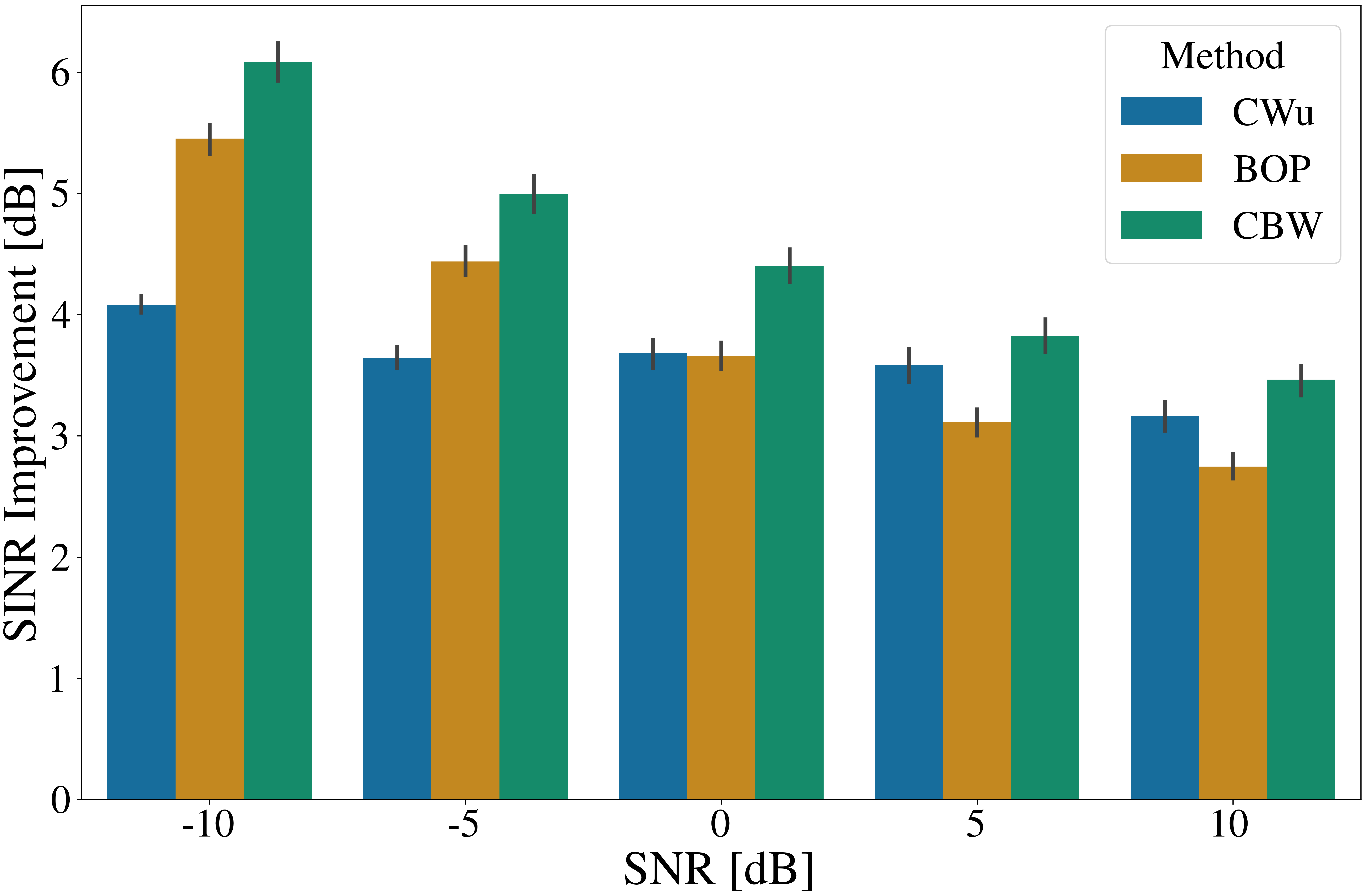}}
  \caption{\gls{SINR} improvement (average and standard deviation) for the considered \gls{RTF} vector estimation methods (\gls{CWu}, \gls{BOP}, \gls{CBW}) for different \glspl{SNR}.}
  \label{fig:SINR_results}
\end{figure}

\subsection{Results}

For different \glspl{SNR}, \autoref{fig:SINR_results} depicts the \gls{SINR} improvement of the \gls{LCMV} beamformer for the considered \gls{RTF} vector estimation methods of the target speaker.
Results were averaged over 72 non-collocated combinations of target and interferer position and 5 different broadband \glspl{SIR} between $\left[\SI{-10}{\dB},\SI{10}{\dB}\right]$ with an increment of $\SI{5}{\dB}$. Results were also averaged over the choice of the reference microphone $m\in\{1,\ldots,M\}$.
For all three methods, the \gls{SINR} improvement increases for lower input \glspl{SNR}, which can be explained by the larger potential improvement in conditions with more noise. For all considered \glspl{SNR}, the \gls{CWu} method yields an \gls{SINR} improvement between \SI{3}{dB} and \SI{4}{dB}. The \gls{BOP} method outperforms the \gls{CWu} method at low input \glspl{SNR}, but yields a lower performance at high input \glspl{SNR}. It can be observed that the proposed \gls{CBW} method outperforms both conventional methods for all input \glspl{SNR}, e.g., achieving an average \gls{SINR} improvement of \SI{6}{\dB} for in the \SI{-10}{\dB} \gls{SNR} condition.

\section{Conclusion}
\label{sec:conclusion}

In this paper, we proposed a \glsfirst{CBW} method to perform \gls{RTF} vector estimation of the second speaker in a dual-speaker scenario with background noise, where the speakers activate successively. In contrast to the conventional methods (\gls{CWu} and \gls{BOP}), the \gls{CBW} method uses an estimate of the \gls{RTF} vector of the first speaker to block the first speaker before noise whitening. When using the estimated \gls{RTF} vectors of both speakers in an \gls{LCMV} beamformer, simulation results demonstrate that the proposed \gls{CBW} method outperforms the conventional \gls{BOP} and covariance whitening methods in terms of \gls{SINR} improvement.
In future work we aim at investigating different combinations of blocking and noise handling, e.g., also introduce noise handling to the \gls{BOP} method.



\bibliographystyle{IEEEtran}
\bibliography{main}
%
%
%
%
%
%
%
%
%

\end{sloppy}
\end{document}